\documentstyle[11pt]{article}
\begin{document}

\begin{center}
THE COUPLING OF THE $f_1(1285)$ MESON TO THE
ISOSCALAR \\AXIAL CURRENT OF THE NUCLEON
\end{center}

\begin{center}
M. Kirchbach
\end{center}
\begin{center}
 Institut f\"ur Kernphysik, TH Darmstadt, D--64289 Darmstadt,
Germany
\end{center}
\begin{center}
and
\end{center}
\begin{center}
D.O. Riska
\end{center}
\begin{center}
Department of Physics, SF--00014 University of Helsinki, Finland
\end{center}

\begin{abstract}
The weak decay as well as
the strong nucleon coupling constants
of the isoscalar axial vector
meson $f_1 (1285)$ are estimated
within the octet quark model.
It is shown that
the empirical value for the coupling of
$\overline{s}s$ quarkonium to nucleons
can be understood by attributing the  $f_1(1285)$-nucleon
coupling to the  $a_0\pi N$ triangle diagram.

\end{abstract}

\section{Introduction}
\setcounter{equation}{0}
One of the most striking empirical indications for the approximate
chiral symmetry
of the strong interaction is the existence of the
two complete low mass vector meson nonets with opposite parities.
An important phenomenological implication of the vector
mesons is their role
in the electromagnetic structure of the baryons.
It was realized early on
that the $\rho$ and $\omega$ mesons can be
used to
parametrize the isovector and isoscalar electromagnetic
formfactors of the nucleon in terms of (di)pole forms
(''vector meson dominance'' VMD) \cite{Na 57} --\cite{Dol 91}.
The VMD idea has also been extended to the
parametrization of the axial isovector
form factor of the nucleon
\cite{Sa 60}.
This is illustrated e.g. by the phenomenological vector meson
Lagrangian proposed  in ref.\cite{Iv 79},
which suggests that the charged weak vector and axial
vector nucleon form factors should be expressed as pole terms that
contain the
masses  $m_\rho$ and
$m_{a_1}$ of the $\rho$ and $a_1$ mesons and their respective couplings
$f_\rho$, and $f_{a_1 }$ to the external $e\overline{\nu}$ weak
current.
We here extend the VMD hypothesis
to the isoscalar axial form factor of the nucleon
along the lines of refs. \cite{Hjo 93}, \cite{Ki 94}
and present a model for the $f_1(1285)NN$ coupling constant.

The paper is organized as follows.
In section 2 we calculate the weak decay couplings of the
$f_1(1285)$ (subsequently denoted by $D$) and
$f_1(1420)$ (subsequently denoted by $E$) mesons and
exploit VMD to derive
relations between hadronic and semileptonic couplings.
In section 3 we calculate
the contribution of the
$D\to (a_0 +\pi) NN$ triangle diagram
to the $DNN$ vertex and demonstrate its
importance for the coupling of the nucleon to an external
isoscalar axial vector current.
The paper ends with a short summary.

\section{The couplings of the $f_1(1285)$ meson to neutral
lepton and nucleon currents}
\setcounter{equation}{0}

The neutral weak axial current of the nucleon
$J_{\mu ,5}^0$ is determined by the
matrix element
of the corresponding isotriplet ($J_{\mu ,5}^{(3)}$), singlet
($J_{\mu ,5}^{(0)}$),
and octet ($J_{\mu ,5}^{(8)}$) quark currents as
\begin{eqnarray}
J_{\mu ,5}^0 & = &
<N\mid (-{1\over 2} J_{\mu , 5}^{(3)}
+{1\over {2\sqrt{6}}} J_{\mu ,5}^{(0)}
-{1\over {2\sqrt{3}}}J_{\mu ,5}^{(8)}\mid N> \, ,\nonumber\\
J_{\mu ,5}^{(3,0,8)} & = &
\overline{\psi}\gamma_\mu\gamma_5 {\lambda^{(3,0,8)}\over 2}\psi ,
\end{eqnarray}
where $\psi$ is the flavour triplet $(u,d,s)^T$ of up, down and
strange quark fields.
The VMD hypothesis implies the
following field--current identities:
\begin{eqnarray}
<N\mid J_{\mu ,5}^{(3,0,8)} \mid N> & = & {m_V^2 \over {2f_V}}V_\mu \, ,
\nonumber\\
\langle \overline{N}N\mid J^V_\mu \mid V\rangle & = &
{m_V^2\over {2f_V}}e^V_\mu ,
\end{eqnarray}
where $V_\mu=(a_1)_\mu$, $(D_0)_\mu$, and $(E_8)_\mu$,
which denote
the fields of the isotriplet, the octet singlet, and the octet
isoscalar axial vector mesons, respectively and $f_V$ are
the corresponding decay constants.
The VMD hypothesis allows derivation of interesting relations
between various hadronic and semileptonic couplings.
For example,
in describing the
$a_1+\pi\to \nu + \overline{\nu}$ annihilation coupling constant
$g_{a_1\pi (\nu\overline{\nu})}$ by means of the chain
$a_1 +\pi \to D\to \nu +\overline{\nu}$ (Fig. 1a)
the following expression is obtained:
\begin{equation}
g_{a_1\pi (\nu \overline{\nu})} =
g_{Da_1\pi}{m_D^2\over {2f_D}}{1\over {m_D^2 - q^2}}  \, .
\end{equation}

Another intriguing relation arises between the pion
weak decay coupling constant $f_\pi$ and
the coupling constant $g_{Da_0 (\mu\overline{\nu}_\mu )} $
of the (hypothetical) weak decay reaction
$D\to a_0 + \mu  +\overline{\nu}_\mu $,
which is of interest for the evaluation
of short range meson exchange currents in nuclear
muon capture reactions
of the type displayed in Fig. 1b,
\begin{equation}
g_{Da_0\pi }{q^2\over {q^2-m_\pi^2}} f_\pi   \,
e^D(x)\cdot J_5^{(\mu\overline{\nu }_\mu)}(x)\,
\stackrel{q^2 \gg m_\pi^2}{\longrightarrow}
g_{Da_0(\mu\overline{\nu}_\mu )}\,  e^D (x)\cdot
 J_5^{(\mu \overline{\nu}_\mu )}(x)\, .
\end{equation}
To obtain this equation we use the Lagrangian density
\begin{equation}
{\cal L}_{Da_0\pi }(x)  =  g_{Da_0\pi } e^D_\lambda (x)
\partial^\lambda
\vec{\pi}(x) \cdot \vec{a_0}(x) \, ,
\end{equation}
In eq. (2.4) $J_5^{(\mu \overline{\nu}_\mu ) }(x)$ is the external weak axial
lepton current.

In the
limit of equal meson masses the nonrelativistic quark model leads
to the following relation
between these decay constants:
\begin{equation}
{1\over f_{a_1}}:{1\over f_{D_0}}:{1\over f_{E_8}}
= 1:-{1 \over \sqrt{6}}:{1 \over  \sqrt{3}}\, .
\end{equation}
The numerical value of $f_{a_1}$
is determined by the second Weinberg sum rule \cite{We 67}
as $f_{a_1}^2 = f_\rho^2$,
while  $f_\rho$ is given by the KSFR relation
as $f_\rho^2 = m_\rho^2/2f_\pi^2$, where $f_\pi$ is the
pion decay constant.
Thus eq. (2.2) allows
expression of the couplings $f_{D_0}$ and $f_{E_8}$
in terms of $f_{a_1}$. This leads to the following
numerical values:
\begin{eqnarray}
f_{D_0} & = &-\sqrt{6}f_{a_1} = - 7.093 \, , \\
f_{E_0} & = & \sqrt{3}f_{a_1}  = 5.016      \, .
\end{eqnarray}
According to the universal
current-current coupling model
the meson-nucleon vertices are
described by effective Lagrangians of the type
\begin{equation}
{\cal L}_{VNN} = f_V\,  V_\eta
\overline{\psi}\gamma^\eta \gamma_5 {\lambda^{(3,0,8) }\over 2 }\psi \, .
\end{equation}

In the quark model the wave functions of the low lying vector mesons are
described as linear combinations of quark--antiquark ($q\overline{q}$)
pairs in (three dimensional) flavour space.
The physical
singlet and isoscalar states of both the vector and axial vector meson
nonets are moreover predicted to be mixed by the so called
''ideal'' mixing angle $\theta_0 = tan^{-1}{1\over \sqrt{2}} $
with the consequence that the states $\overline{s}s$
and $(\overline{u}u +\overline{d}d)/\sqrt{2}$ decouple
completely.
This is equivalent to  a suppression of
$\overline{s}s\to (\overline{u}u+\overline{d}d)/\sqrt{2}$ transitions
(the Okubo--Zweig--Iizuka rule).
Consequently the decay of, say, the $\phi (\overline{s}s)$
meson into three pions
via the chain
$\phi\to \rho\pi\to \pi^+\pi^-\pi^0$ is represented in QCD by
nonplanar diagrams, which are known to be suppressed relative
to the leading planar graphs by  $1/N_c$.
The constants
$f_D$ and $f_E$ corresponding to
the ideally mixed $D$ and $E$ mesons
are then given as
\begin{eqnarray}
f_D & = &
f_{D_0} \, cos\theta_0 -f_{E_8}\, sin\theta_0 = -3f_{a_1}  \, ,\\
f_E  &= & f_{D_0}  \, sin\theta_0 +
f_{E_8} \, cos\theta_0 \, = 0.
\end{eqnarray}
In this way the couplings of the isoscalar axial vector mesons to the
external lepton current equal (up to the sign)
the corresponding couplings
of the isocalar vector mesons for which the following
relations have been obtained within the quark model \cite{deSwa 75},
$f_\omega = 3f_{a_1}$, and $f_\phi =0$.
The experimental values for the physical mixing angle
(subsequently denoted by $\theta$), which have been extracted
from the empirical meson masses with
linear or quadratic mass formulae, disagree  however with
the ideal one. This difference is comparatively small
in the case of the vector meson nonet (about $5^o$).
Even so, the small observed $\omega -\phi$ mixing has a non--negligible
impact on the weak vector strangeness radius of the nucleon--
a result obtained in \cite{Ja 89}, \cite{Fo 93} under the assumption
of VMD of the isoscalar weak vector current.
The axial vector meson nonet is clearly also non-ideal, as the
empirical mixing value for it is
$\theta =\theta_0 +\epsilon  =50^o\pm 2.7^o\pm 3.6^o$  \cite{Bo 92}.

The wave functions of the isoscalar axial vector mesons
that take into account the mixing between the strange and non strange
quarkonia are

\begin{eqnarray}
D(1285)\equiv f_1(1285) & = &
cos \epsilon {{\overline{u}u + \overline{d}d}\over {\sqrt{2}}}
-sin\epsilon \,\, \overline{s}s
\nonumber\\
E(1420)\equiv f_1(1420) & = &
sin\epsilon {{\overline{u}u  + \overline{d}d}\over {\sqrt{2}}}
+cos \epsilon\,\,  \overline{s}s \, .
\end{eqnarray}

The substantial deviation of the D--E mixing from the ideal mixing
angle is explained qualitatively in QCD as being due to
strong nonperturbative effects in the
isoscalar axial channel \cite{Shi 79}.
In the case of the mixed isoscalar axial vector mesons eqs. (2.9)--(2.10)
are replaced by
\begin{eqnarray}
\hat{f}_D & = &  f_D cos\epsilon -
f_E sin\epsilon   =  -3f_{a_1} cos \epsilon \, , \\
\hat{f}_E & = &  f_D sin \epsilon +
f_E cos\epsilon  = -3f_{a_1}\sin\epsilon  \,  .
\end{eqnarray}

The isoscalar axial vector mesons D(1285) and E(1420)
are expected to be important as intermediate states
in $N\overline{N}$ annihilation as well as in neutrino scattering
processes.
In this  context the weak decay couplings
of the $f_1(1285)/f_1(1420)$ mesons can be associated
with the
quantities $\hat{f}_D/\hat{f}_E$.
The interpretation of the latter as
the strong meson--nucleon couplings is, however, not obvious.
This follows from the observation that
eq. (2.1)  in fact reduces to
\begin{eqnarray}
J_{\mu ,5}^0 & = & -{1\over 2}<N\mid J_{\mu ,5}^{(3)}\mid N> +
               {1\over 4}  <N\mid \overline{s}\gamma_\mu\gamma_5 s
\mid N>\, ,\nonumber\\
& = & -{g_A\over 2} \overline{u}_N\gamma_\mu\gamma_5 {\tau_3\over 2}u_N
+ {1\over 2}G_1^s \overline{u}_N\gamma_\mu\gamma_5u_N \, ,
\end{eqnarray}
where $g_A$ is the weak proton axial coupling constant, and
$G_1^s$ denotes the so called weak isoscalar nucleon coupling
($G_1^s =- 0.13 \pm 0.04 $ \cite{Ell 93}).
This equation suggests that
the coupling constants
in the $f_1(1285)NN$ and $f_1(1420)NN$ vertices
(in turn denoted by $g_{DNN}$, and $g_{ENN}$) will
be determined
by the $\overline{s}s$ components of the corresponding
wave functions
according to
\begin{eqnarray}
g_{DNN} &  = & g_{(\overline{s}s)} sin \epsilon \, , \nonumber\\
g_{ENN} & = &  g_{(\overline{s}s)} cos \epsilon  \, .
\end{eqnarray}
Here $g_{(\overline{s}s)} $ denotes the coupling of the
purely strange quarkonia to the nucleon.
By the parity invariance (chiral symmetry) of the strong interaction
one can
assume that the value  $g_{(\overline{s}s)}$
from the axial vector meson
nonet equals the empirical
value of  $g_{(\overline{s}s)} = 6.75$,
which is extracted from data
on the vector meson nonet \cite{Ge 76}.
This yields the following relation
\begin{equation}
\mid g_{DNN}\mid  = \sin\epsilon g_{(\overline{s}s)} = 1.72 \, .
\end{equation}

In the following section we show that
by attributing
the f$_1$(1285)NN coupling to the
$a_0\pi N$ triangle diagram, which is
the dominant one loop contribution
to the isoscalar weak axial nucleon coupling,
the empirical value
for the coupling of the $\overline{s}s$ quarkonium
to the nucleon is well understood.

\section{The $\pi a_0 N$ triangle diagram contribution
to $g_{DNN}$}
\setcounter{equation}{0}

The $\pi a_0 N $ triangle diagram contribution to the
$D$-nucleon coupling $g_{DNN}$ (Fig. 2) and the corresponding form factor
can be calculated using the $D a_0 \pi$ Lagrangian (2.5)
and the following  Lagrangians for the
$\pi NN$ and $a_0 NN$ couplings

\begin{eqnarray}
{\cal L}_{\pi NN} & = & \frac{f_{\pi NN}}{m_\pi}\bar\psi\gamma_5\gamma^\mu
\partial_\mu \vec\phi\cdot\vec \tau\psi, \\
{\cal L}_{a_0NN}& = & g_{a_0 NN}\bar\psi\vec\phi\cdot\vec\tau\psi \, .
\end{eqnarray}
Here $f_{\pi NN}$ and $g_{a_0 NN}$
in turn denote the pseudovector $\pi$NN
and  the scalar  $a_0NN$ coupling constants, for which we
adopt the values $f_{\pi NN}^2/4\pi = 0.08$ and
$g_{a_0 NN}^2/4\,\pi=1.075$, respectively. These values are implied
by the semiphenomenological Bonn one boson
exchange model for the nucleon-nucleon interaction
\cite{Els 87}. We define the effective D-nucleon vertex by
the scalar product between the weak isoscalar axial vector current
of the nucleon ($J_{\mu ,5}^0$) and the $f_1(1285)$ field (denoted
by $V_D^\mu$) as
\begin{equation}
J^0_5 \cdot V_D =  iG_1^s \bar u(\vec{p\, })\gamma_5\gamma^\mu u(\vec{p\, })
\, f_{DNN} e^D_\mu  .
\end{equation}
The product $f_{DNN}G_1^s$ equals now the quantity $g_{DNN}$
introduced by eq. (2.16)
with $g_{DNN}=g_{DNN}(-m_D^2)$. To regularize the integral
in the triangle diagrams in Fig.2 we introduce the same monopole
formfactors at the $\pi$NN and $a_0$NN  vertices as in
the potential model \cite {Els 87}. The two triangle
diagrams then give the following contribution to the
DNN vertex:
\begin{equation}
g_{DNN} (q^2)=\frac{3}{16\pi^2}g_{\pi NN}g_{Da_0\pi}g_{a_0 NN}
\int_0^1dx\int_0^1dy x log
\lbrace \frac{Z(m_\pi,\Lambda_a)Z(m_a,\Lambda_\pi)}
{Z(m_\pi,m_a)Z(\Lambda_\pi,\Lambda_a)}\rbrace.
\end{equation}
Here $\Lambda_\pi$ and $\Lambda_a$ are the mass parameters
in the monopole vertex factors, for which we use
the values
1.3 GeV and 2.0 GeV respectively, and
the function $Z(m_1,m_2)$ is
defined as
\begin{equation}
Z(m_1,m_2)=-q^2xy^2(x-1)+m_N^2x^2(1-y)^2+(1-x)m_1^2+xym_2^2.
\end{equation}
The DNN coupling constant is obtained by setting
$q^2=-m_D^2$ in eq. (3.4).
Note that as $m_D> m_\pi+m_{a_0}$ the
expression (3.4) has a small imaginary part when the D-meson is
on shell, which we shall ignore.
Using for $g_{Da_0\pi}$ the value 5.7  from the experimental
partial decay width \cite{PARTD},
the real part of (3.4) at $q^2=-m_D^2$ is then 1.9, which we shall
interpret as the value for the DNN coupling constant, i.e.
\begin{equation}
\mid g_{DNN}\mid  = 1.9 \, .
\end{equation}
The
$q^2$ dependence of $g_{DNN}$ is shown in Fig.3. The $q^2$-dependence
corresponds approximately to that of a monopole form with a
mass scale parameter $\Lambda\sim$ 2 GeV.\\

\section{Results and discussion}
\setcounter{equation}{0}

It has long been known that the $\phi$NN
coupling is strongly underpredicted in
the quark model as seen from the unexpectedly large
value for the
$(\overline{s}s)$NN coupling extracted from data analyses in
\cite{Ge 76}.
Comparison
of our result from eq. (3.6) shows
that the numerical value for $g_{DNN}$
associated with the $a_0\pi N$ triangle diagram agrees fairly well
with the empirical estimate given in eq. (2.17).
Consequently, in spite of the small decay width of the $f_1(1285)$
meson the coupling constant in the $f_1(1285)NN$ vertex
seems to be completely exhausted by the contribution of the $a_0\pi N$
triangle diagram.\\

The result that the empirical value for that coupling constant
can be explained by the $\pi a_0$ triangle diagram alone
shows that the meson model is able to predict
realistic results for reactions that involve
the coupling of the $s\bar s$ system to nucleons.\\

Using the experimental value of $G_1^s$ from \cite{Ell 93}
the value of $f_{DNN}$ is obtained as $f_{DNN} = -14.61$
which is compatible with the size of
$\hat{f}_D =-3f_{a_1}cos\epsilon = -17.2 $ obtained by means
of the KSFR relation.
This suggests that the
vector meson dominance model may be
applied in the case of isoscalar axial vector currents.
\\

\vspace{1cm}
\noindent
{\bf Acknowledgements}
\vspace{0.5cm}

\noindent
M.K. is grateful to the hospitality of the Department of Physics of
the University of Helsinki and the Academy of Finland for travel
support.

\newpage
\noindent
{\bf Figure captions}
\vspace{0.5cm}

Fig. 1 \hspace{0.2cm} (a)
$D$ exchange contribution to $a_1\pi \rightarrow
\nu\bar \nu$,
(b)$\pi$ exchange contribution to $D\rightarrow a_0\mu\bar \nu$.\\

Fig. 2 \hspace{0.2cm} The $\pi a_0$ triangle diagram contribution to
the $D$-nucleon coupling.\\

Fig. 3 \hspace{0.2cm} The momentum dependence of the $\pi a_0$
triangle diagram contribution to the $DNN$ coupling $g_{DNN}(q^2)$.
The units of $q^2$ are $GeV^2$.

\end {document}